# Up-conversion single-photon detectors based on integrated periodically poled lithium niobate waveguides

Fei Ma,[1] Long-Yue Liang,[2] Jiu-Peng Chen,[1] Yang Gao,[3] Ming-Yang Zheng,[3] Xiu-Ping Xie,[3,4] Hong Liu,[2] Qiang Zhang,[1,4,*] and Jian-Wei Pan[1,*]

[1]*Hefei National Laboratory for Physical Sciences at Microscale and Department of Modern Physics, University of Science and Technology of China, Hefei, Anhui 230026, China*
[2]*State Key Laboratory of Crystal Materials, Shandong University, Jinan, Shandong, 250100, China*
[3]*Shandong Institute of Quantum Science and Technology Co., Ltd., Jinan, Shandong 250101, China*
[4]*Jinan Institute of Quantum Technology, Jinan, Shandong 250101, China*
qiangzh@ustc.edu.cn, pan@ustc.edu.cn

## 1. Abstract

We demonstrate up-conversion single-photon detectors based on integrated periodically poled lithium niobate waveguides, which incorporate two mode filters and a directional coupler. The two mode filters are optimized for the fiber-waveguide coupling efficiencies for 1550 nm and 1950 nm respectively while the directional coupler plays the role of wavelength combiner, making the overall system portable and low-cost. The two wavelengths pump each other in our detection system. We achieve detection efficiencies of 28% for 1550 nm and 27% for 1950 nm, respectively. This scheme provides an efficient integrated single-photon detection method for any two well-separated spectral bands in the whole low-loss range of lithium niobate waveguides.

## 2. Introduction

Up-conversion single-photon detector (SPD) [1-8] is a promising technique to extend the photon-counting capability of well-developed silicon avalanche photodiode (APD) to the infrared region. When a weak signal mixes with a strong pump via sum-frequency generation (SFG) in periodically poled lithium niobate (PPLN) waveguides, infrared-band signal photons can be efficiently converted into the near-infrared or visible window, and then be efficiently detected by a silicon APD. This detection method is an attractive alternative to direct detection by InGaAs/InP APD which usually suffers from relatively low efficiency [9] and saturation counting rate [10], and high-performance superconducting SPD which requires cryogenic cooling and exquisite temperature control may increase the complexity of the detection system [9, 11-13]. Nowadays, up-conversion SPDs have been widely applied in quantum key distribution (QKD) systems [14-17].

Previous studies [8] show that an up-conversion SPD with a single waveguide can detect single photon at either of the two input wavelengths (1.3 µm and 1.55 µm) by exchanging the roles of the signal and the pump. However this only works when the pump and the signal wavelengths are not separated far away [8, 18, 19]. Recently, PPLN waveguides are used for a high-performance up-conversion detector based on the long-wavelength pumping technology which involves two wavelengths of significant frequency difference [6, 20, 21]. It is difficult to optimize the coupling efficiencies of both the signal and the pump with a single mode filter when both waves must be coupled into the fundamental waveguide mode. For the reported PPLN-waveguide-based up-conversion SPDs, the mode filter integrated in the waveguides was optimized for the fiber-waveguide coupling efficiency of the 1550 nm signal only, at the expense of low coupling efficiency for the longer pump wavelength. Therefore, as an example, 2 μm detection using an up-conversion SPD with coupling efficiency optimized for 1550 nm band suffers from low detection efficiency because 2 μm is now the signal [22]. Moreover, a new waveguide design with high coupling efficiency for both the signal and pump is needed to lower the requirement for pump power and thus reduce the system cost.

In this letter we demonstrate an integrated waveguide with a special configuration [23, 24], including two individual mode filters designed to optimize the fiber-waveguide coupling efficiencies for the two SFG wavelengths respectively, which are well separated in the low-loss range (0.5-2.6 μm) of typical reverse-proton-exchange (RPE) lithium niobate waveguides. A directional coupler playing the role of wavelength combiner is integrated into the waveguide structure to couple both waves into the frequency conversion region and realize efficient up-conversion.

Here we choose 1.55 μm band and 2 μm band to perform the single-photon detection, not only because of their great signal-pump frequency difference, but also for their extensive applications. SPDs operating at 1.55 μm telecom band are the core components for optical fiber quantum communication systems [25]. Moreover, 1.55 μm band is also widely applied in the fields of single-photon-level spectrometer [21] and optical time domain reflectometry (OTDR) [26]. On the other hand, 2 μm matches one absorption band of $CO_2$ [27], meaning that 2 μm waves can be used for the study of global warming. Furthermore, the ground testing of a 2 μm Doppler aerosol wind Lidar system has been demonstrated [28]. In addition, the lowest propagation losses for hollow-core photonic bandgap fibers occur near 2 μm band, which can provide a 100-fold enhancement of the overall capacity of broadband core networks [29].

With the integrated waveguide structure, we demonstrate up-conversion SPDs by swapping the roles of the input waves as signal and pump without sacrificing the fiber-waveguide coupling efficiency of any wave. Detection efficiencies (DE) of 28% at 1550 nm and 27% at 1950 nm are achieved, with noise count rates (NCR) of 150 counts per second (cps) and 24300 cps respectively.

## 3. Design and Optimization of the Integrated PPLN Waveguides

We fabricate the integrated reverse-proton-exchange PPLN waveguides [30, 31] for SFG of 1550 nm and 1950 nm with a total length of 60 mm. Congruent lithium niobate crystal is used. The integrated waveguide structure is shown in Fig. 1, which consists of a bent waveguide and a straight waveguide with a center-center separation of 126 μm at the entrance. The main features of the integrated structure include two individual mode filters optimized for the fiber-waveguide coupling efficiencies of 1550 nm and 1950 nm respectively, adiabatic tapers, S-bends, a directional coupler working as a wavelength combiner, and a uniform straight waveguide with 48-mm-long quasi-phase-matching (QPM) gratings for optical frequency mixing, which is poled with a period of 20 μm.

We fix a RPE dose of 0.78 μm [32] for the lowest waveguide loss and fabricate a series of waveguides with different mode filter widths varying from 2 μm to 7 μm. A polarization maintaining (PM) 1550 nm fiber pigtail is employed to measure fiber-waveguide coupling efficiencies for 1550 nm and 1950 nm. As shown in Fig. 2(a) and Fig. 2(b), the optimal mode filter widths for 1550 nm and 1950 nm are 3.5 μm and 5.0 μm, with the highest fiber-waveguide coupling efficiencies of ~80% and 75%, respectively.

Here we present the simulation results of fiber-waveguide coupling efficiencies versus different mode filter widths. As shown in Fig. 2(c), the optimal mode filter widths for 1550 nm and 1950 nm are ~2.7 μm and 3.7 μm, with the highest fiber-waveguide coupling efficiencies of ~ 97% and 92%, respectively. We observe quantitative discrepancies between experimental results in Fig. 2(a), Fig. 2(b) and simulation results in Fig. 2(c). There are two main reasons. On the one hand, the nonlinear diffusion waveguide model is based on refractive index of the planar waveguides, with extra undercutting parameters considered for wide waveguides to match nonlinear experimental results at 1550 nm [32], which is therefore not very suitable for exact mode filter simulation with narrow width. On the other hand, the imperfect fabrication in practice is always inevitable and the actual coupling efficiencies cannot match the values given by simulation. However, both experimental and simulation results show the same tendency, and the simulation results can tell us when we should employ two mode filters instead of one. For example, as indicated with the vertical dashed line in Fig. 2(b), when the coupling efficiency for 1550 nm reaches the highest point with a 2.7 μm wide mode filter, the coupling efficiency for 1750 nm, 1950 nm and 2150 nm waves are 5%, 15% and 25% lower, respectively. For 1550 nm and 1750 nm, the two adjacent wavelengths can only be separated with a long directional coupler, while the gain in fiber-waveguide coupling efficiency from using two mode filters is too small to compensate for the loss caused by the long directional coupler. For 1550 nm and 1950 nm or 2150 nm, the two wavelengths are further apart and the gain in coupling efficiency from using two mode filters is much higher than the loss caused by the directional coupler. Experimental data below would confirm this point. Therefore, it is reasonable to use two mode filters when the two SFG wavelengths around 1550 nm differ by ≥400 nm, such as we demonstrate in this letter.

Generally, such a scheme is applicable for SFG of any two spectral bands in the transparency range of lithium niobate (0.4-5 μm) with large wavelength difference. However, we need to further consider the propagation loss in waveguides. Figure 2(d) is the transmission spectrum of a typical proton exchanged lithium niobate chip, showing strong absorption below 500 nm and around 2.86 μm [32, 33]. Moreover, the cutoff wavelength is ~2.6 μm for typical RPE waveguides used for SFG with a dose of 0.78 μm [32]. Therefore, the low-loss wavelength range for typical reverse-proton-exchange (RPE) lithium niobate waveguides is 0.5-2.6 μm, for which our scheme of up-conversion SPDs applies. Of course, we may use MgO:PPLN waveguides instead to avoid photo-refractive damage when high power at wavelengths <1 μm is involved.

To guide the 1950 nm wave to the directional coupler, we fabricate S-bend of cosine type which has the minimum bending loss [34-36]. As shown in Fig. 3(a), the bending loss is a decreasing function of L, the length of the S-bend. The bending loss drops to ~0 for L≥3.5 mm. Therefore we use L=3.5 mm to enable the longest QPM gratings in the device with limited total length.

A symmetrical directional coupler is designed so that the 1950 nm wave is coupled to the adjacent waveguide while the 1550 nm wave is kept in its own path after they propagate a characteristic distance, Lc, which can be predicted from coupled mode theory [37, 38]. Based on simulation results, the waveguide width and edge-edge spacing are fixed at 5.5 μm and 8 μm, respectively. A typical measurement of Lc involves fabricating a set of waveguides with different coupler lengths, and measuring the coupling efficiency for each length. The 1950 nm coupling efficiency is defined as the ratio between the output power of the straight waveguide and the total output power for 1950 nm; the 1550 nm coupling loss is defined as the ratio between the output power of the bent waveguide and the total output power for 1550 nm. As shown in Fig. 3(b), when the coupler length is Lc=4.8 mm, the 1950 nm coupling efficiency is >95%, and the 1550 nm coupling loss is ~0.5 dB. The latter is close to the insertion loss of a conventional wavelength division multiplex (WDM) device and is acceptable. For a counter example, we consider the combination of 1550 nm and 1750 nm. According to the simulation results, the optimal coupler length is 18.6 mm, about 3.8 times longer than that for the combination of 1550 nm and 1950 nm, and will reduce the QPM gratings length in a device with fixed total length. Moreover, the coupling loss of 1550 nm will now increase to ~1.5 dB, so it is not worth to use the directional coupler for such a situation.

We fabricate the integrated PPLN waveguides based on above test results. The directional coupler length is 4.8 mm, the S-bend length is 3.5 mm, and the mode filter widths for 1550 nm and 1950 nm are 3.5 µm and 5 µm respectively. The input and output facets of the waveguides are anti-reflection (AR) coated for all wavelengths of interest to eliminate the Fresnel reflection loss. We use two PM 1550 nm fibers that are terminated in a silicon V-groove array with a core spacing of 126 µm, i.e. a PM fiber array (FA), for fiber pigtailing at the input port of the waveguides.

With these parameters the throughputs of 1550 nm and 1950 nm at the arm of straight waveguide are measured to be 62% and 61% respectively, which are consistent with our estimations by combining the losses of each waveguide components, as shown in Table 1.

Finally, we characterize the phase matching of the integrated waveguides. For 1550 nm up-conversion process pumping by 1950 nm, the SFG tuning curve [6, 20] is shown in Fig. 4. We fix the pump laser wavelength at 1950 nm and sweep the signal laser around 1.55 µm band. The central phase-matching wavelength is 1548.15 nm at room temperature with a full width at half maximum (FWHM) of 0.63 nm, which agrees well with the numerical simulation result. It is the same case for 1950 nm up-conversion process pumping by 1550 nm, since the signal wavelength is not tunable, the corresponding FWHM can be calculated as ~0.92 nm by numerical simulation.

## 4. Characterization and Performance of Up-conversion Single-photon Detectors based on the Integrated PPLN Waveguides

A schematic of the experimental setup for up-conversion SPDs based on the integrated PPLN waveguides is shown in Fig. 5.

For 1550 nm up-conversion SPD, as shown in Fig. 5(a), a continuous-wave single-frequency PM fiber laser amplified by a thulium-doped fiber amplifier (TDFA) is used as the pump source. Single-photon-level 1550 nm signal with a power of -98.92 dBm, i.e. one-million photons per second at the input port of the waveguides is produced with a TSL-510 tunable semiconductor laser, a polarizer, and two PM 1550 nm variable optical attenuators. A 1/99 1550 nm PM beam splitter and a calibrated power meter are exploited to monitor the input signal power [39]. Employing PM-fiber components for both the pump and the signal improves the stability of the whole system.

For 1950 nm up-conversion SPD, as shown in Fig. 5(b), the 1950 nm signal, which is one-million photons per second at the input port of the waveguides with a power of -99.92 dBm, is generated with a PM fiber laser together with five 1/99 1950 nm PM beam splitters in series. A calibrated power meter is used to monitor the input signal power. A TSL-510 single-frequency tunable semiconductor laser serves as the pump seed, whose output wavelength is fixed at 1550 nm. The seed is then amplified with an erbium-doped fiber amplifier (EDFA) which produces a maximum power of 200 mW at 1550 nm. Unwanted photons at 1950 nm and 863 nm generated in the EDFA are removed using a 1550 nm/1950 nm WDM and a 1550 nm/863 nm WDM. As RPE waveguides support only TM-polarized modes, polarization controllers are used to adjust the polarization of the 1550 nm wave.

The output wavelengths of 1.55 μm and 2 μm lasers are fixed at 1550 nm and 1950 nm respectively. A temperature-controlled oven keeps the waveguides working at an optimal temperature to maintain the phase-matching condition.

As shown in Fig. 5(c), the up-conversion photons generated from SFG process in the integrated PPLN waveguides and the remnant pump are collected with an AR-coated aspheric lens. The remnant pump is then removed with a dichroic mirror (DM) and the up-conversion photons are reflected with reflectance above 99.5%. A 785 nm long-pass filter, a 945 nm short-pass filter and a band-pass filter centered at 857 nm with a bandwidth of 30 nm are used in combination to block the noises coming from the strong pump, including spontaneous Raman scattering (SRS) noise, parasitic noise caused by imperfect periodic poling structure, spontaneous parametric down-conversion (SPDC) noise when 1550 nm acting as the pump, and second and third harmonic generation photons [20]. The transmittance of long-pass filter, short-pass filter and band-pass filter are all measured above 99%. Then, two etalons with the spectral FWHM of 0.1 nm and free spectral range (FSR) of 0.3 nm, and other two etalons with FWHM of 0.1 nm and FSR of 0.5 nm are employed to further reduce the SRS noise [19]. The effective pass bandwidth and overall transmittance of the etalons stack are measured as 0.07 nm and 93.2%, respectively. Finally, an 863 nm collimator and a multimode fiber are employed to collect the SFG photons with coupling efficiency of more than 99.9%, which are then detected by a silicon APD. We achieve a free-space filtering system with a very low loss of 0.46 dB, which benefits from two aspects as following. For one thing, the aspheric lens, the collimator and the end face of multimode fiber are all AR coated for 863 nm, the transmission losses are negligible. For another, a series of optimal optical components are selected in our experiment.

Since the FWHM bandwidth of the filtering system for SFG photons is 0.07 nm, the detectable signal bandwidth of our detectors at 1550 nm and 1950 nm are filtered into 0.23 nm and 0.36 nm, respectively.

The DE, NCR and signal-to-noise ratio (SNR) for our up-conversion SPDs are recorded by gradually decreasing the pump power, and are shown in Fig. 6. The DE is obtained by dividing the number of detected count rate by one million after NCR subtraction, and the NCR contains the silicon APD's intrinsic noise of 50 cps. SNR is the ratio of detected count rate after NCR subtraction to NCR, which is a convenient metric to characterize noise performance, as shown in Fig. 6(b) and Fig. 6(d). At the maximum conversion point, we achieve a DE of 28% at 1550 nm with a NCR of 150 cps. The optimal phase-matching temperature is 33 °C.

For 1950 nm single-photon detection, the optimal phase-matching temperature is 37 °C, and we achieve a DE of 27% with a NCR of 24300 cps at the maximum pump power of 188 mW. The increment of phase-matching temperature may be caused by the green-light induced photo-refractive effect which is originated from the third harmonic generation of 1550 nm pump.

We note the optimal phase-matching temperature decreases monotonically from 37 °C to 33 °C with decreasing 1550 nm pump power. Since the 1550 nm pump power is slightly insufficient as shown in Fig. 6(c), we make a sine-squared fitting [8, 38] and the results show that the maximal conversion has been achieved basically.

In combination with the total waveguide losses of 2.05 dB for 1550 nm and 2.12 dB for 1950 nm, the free space loss of 0.46 dB and the silicon APD's DE of ~50%, our experimental results are consistent with the estimated DE. The difference of NCR for the two cases can be explained by the larger gain for Stokes scattering (1550 nm pump) as compared to anti-Stokes scattering (1950 nm pump) [20]. Moreover, 1550 nm short-wavelength pumping will introduce a large amount of SPDC noise while 1950 nm long-wavelength pumping can eliminate it [20].

For our integrated PPLN waveguides-based 1550 nm up-conversion SPD, the required 1950 nm pump power has been reduced to ~46% of that required in the reported up-conversion SPD [22], attributed to mode-filter optimization, save of WDM and AR-coating for both signal and pump. Owing to the high throughput of 1950 nm signal, we achieve a DE of up to 27% for the 1950 nm up-conversion SPD. Our up-conversion SPDs have two more advantages as following: firstly, saving the external wavelength combiner makes the system portable and low-cost; secondly, the filtering system using dielectric filters only can be integrated as a fiber filter, which could make the overall system more stable and practical in field [39].

## 5. Conclusion

We design and fabricate the integrated PPLN waveguides, incorporated with two mode filters optimized for the fiber-waveguide coupling efficiencies for 1550 nm and 1950 nm respectively, and an integrated structure with optimal directional coupler length and S-bend length. Based on the integrated PPLN waveguides, we build up-conversion SPDs and achieve detection efficiencies of 28% at 1550 nm and 27% at 1950 nm, with noise count rates of 150 cps and 24300 cps respectively. Employing the directional coupler in up-conversion SPD technology makes the system portable and low-cost as the external wavelength combiner is saved. Apart from QKD, an immediate application will be detection of $CO_2$ [27], which has two absorption bands at 1.5 μm and 2 μm. Moreover, our design shows an approach to lower the demand for pump power, which might be of great interest for particular applications using special wavelengths. This scheme can cover any two spectral bands with significant wavelength differences in the low-loss range (0.5-2.6 μm) of typical RPE lithium niobate waveguides. Last but not least, our SPDs can be integrated as an all-fiber system, which is a promising robust counter and useful in many fields.

## Acknowledgments
The authors thank Dai-Ying Wei for the packaging of the waveguides, Quan Yao and Yi Tao for the help in the fabrication process.

## References

1. P. Kumar, "Quantum frequency conversion," Opt. Lett. **15**, 1476-1478 (1990).

2. A. P. Vandevender and P. G. Kwiat, "High efficiency single photon detection via frequency up-conversion," J. Mod. Opt. **51**, 1433-1445 (2004).

3. M. A. Albota and F. N. C. Wong, "Efficient single-photon counting at 1.55 μm by means of frequency up-conversion," Opt. Lett. **29**, 1449-1451 (2004).

4. R. T. Thew, H. Zbinden, and N. Gisin, "Tunable up-conversion photon detector," Appl. Phys. Lett. **93**, 071104 (2008).

5. L. Ma, O. Slattery, and X. Tang, "Single photon frequency up-conversion and its applications," Phys. Rep. **521**, 69-94 (2012).

6. H. Kamada, M. Asobe, T. Honjo, H. Takesue, Y. Tokura, Y. Nishida, O. Tadanaga, and H. Miyazawa, "Efficient and low-noise single-photon detection in 1550 nm communication band by frequency up-conversion in periodically poled $LiNbO_3$ waveguides," Opt. Lett. **33**, 639-641 (2008).

7. Z. Xie, K. Luo, H. Herrmann, C. Silberhorn, and C. W. Wong, "Efficient single-photon frequency up-conversion at telecommunication wavelengths with Ti-indiffused periodically-poled $LiNbO_3$ waveguides," in *Conference on Lasers and Electro-Optics*, (Optical Society of America, 2015), paper FM3A.3.

8. C. Langrock, E. Diamanti, R. V. Roussev, Y. Yamamoto, M.M. Fejer, and H. Takesue, "Highly efficient single-photon detection at communication wavelengths by use of up-conversion in reverse-proton-exchanged periodically poled $LiNbO_3$ waveguides," Opt. Lett. **30**, 1725-1727 (2005).

9. M. D. Eisaman, J. Fan, A. Migdall, and S. V. Polyakov, "Invited review article: Single-photon sources and detectors," Rev. Sci. Instrum. **82**, 071101 (2011).

10. B. Korzh, N. Walenta, T. Lunghi, N. Gisin, and H. Zbinden, "Free-running InGaAs single photon detector with 1 dark count per second at 10% efficiency," Appl. Phys. Lett. **104**, 081108 (2014).

11. C. M. Natarajan, M. G. Tanner, and R. H. Hadfield, "Superconducting nanowire single-photon detectors: physics and applications," Supercond. Sci. Technol. **25**, 063001 (2012).

12. C. Schuck, W. H. P. Pernice, and H. X. Tang, "NbTiN superconducting nanowire detectors for visible and telecom wavelengths single photon counting on $Si_3N_4$ photonic circuits," Appl. Phys. Lett. **102**, 051101 (2013).

13. I. E. Zadeh, J. W. N. Los, R. B. M. Gourgues, V. Steinmetz, G. Bulgarini, S. M. Dobrovolskiy, V. Zwiller, and S. N. Dorenbos, "Single-photon detectors combining high efficiency, high detection rates, and ultra-high timing resolution," APL Photonics **2**, 111301 (2017).

14. R. T. Thew, S. Tanzilli, L. Krainer, S. C. Zeller, A. Rochas, I. Rech, S. Cova, H. Zbinden, and N. Gisin, "Low jitter up-conversion detectors for telecom wavelength GHz QKD," New J. Phys. 8, 32-43 (2006).

15. Eleni Diamanti, Hiroki Takesue, Carsten Langrock, M. M. Fejer, and Yoshihisa Yamamoto, "100 km differential phase shift quantum key distribution experiment with low jitter up-conversion detectors," Opt. Express **14**, 13073-13082 (2006).

16. H. Xu, L. Ma, A. Mink, B. Hershman, and X. Tang, "1310-nm quantum key distribution system with upconversion pump wavelength at 1550 nm," Opt. Express 15, 7247-7260 (2007).

17. Y. Liu, T. Chen, L. Wang, H. Liang, G. Shentu, J. Wang, K. Cui, H. Yin, N. Liu, L. Li, X. Ma, J. S. Pelc, M. M. Fejer, C. Peng, Q. Zhang, and J. Pan, "Experimental Measurement-Device-Independent Quantum Key Distribution," Phys. Rev. Lett. 111, 130502 (2013).

18. J. S. Pelc, P. S. Kuo, O. Slattery, L. Ma, X. Tang, and M. M. Fejer, "Dual-channel, single-photon up-conversion detector at 1.3 μm," Opt. Express **20**, 19075-19087 (2012).

19. P. S. Kuo, J. S. Pelc, O. Slattery, Y.-S. Kim, M. M. Fejer, and X. Tang, "Reducing noise in single-photon-level frequency conversion," Opt. Lett. **38**, 1310-1312 (2013).

20. J. S. Pelc, L. Ma, C. R. Phillips, Q. Zhang, C. Langrock, O. Slattery, X. Tang, and M. M. Fejer, "Long-wavelength-pumped up-conversion single-photon detector at 1550 nm: performance and noise analysis," Opt. Express **19**, 21445-21456 (2011).

21. G. Shentu, J. S. Pelc, X. Wang, Q. Sun, M. Zheng, M. M. Fejer, Q. Zhang, and J. Pan, "Ultralow noise up-conversion detector and spectrometer for the telecom band," Opt. Express **21**, 13986-13991 (2013).

22. G. Shentu, X. Xia, Q. Sun, J. S. Pelc, M. M. Fejer, Q. Zhang, and J. Pan, "Up-conversion detection near 2 μm at the single photon level," Opt. Lett. **38**, 4985-4987 (2013).

23. M. H. Chou, J. Hauden, M. A. Arbore, and M. M. Fejer, "1.5-μm-band wavelength conversion based on difference-frequency generation in $LiNbO_3$ waveguides with integrated coupling structures," Opt. Lett. **23**, 1004-1006 (1998).

24. J. S. Pelc, L. Yu, K. De Greve, P. L. McMahon, C. M. Natarajan, V. Esfandyarpour, S. Maier, C. Schneider, M. Kamp, S. Höfling,

R. H. Hadfield, A. Forchel, Y. Yamamoto, and M. M. Fejer, "Down-conversion quantum interface for a single quantum dot spin and 1550-nm single-photon channel," Opt. Express **20**, 27510-27519 (2012).

25. R. H. Hadfield, "Single-photon detectors for optical quantum information applications," Nat. Photonics **3**, 696–705 (2009).
26. M. Legré, R. Thew, H. Zbinden, and N. Gisin, "High resolution optical time domain reflectometer based on 1.55 μm up-conversion photon-counting module," Opt. Express **15**, 8237-8242 (2007).
27. G. Ehret, C. Kiemle, M. Wirth, A. Amediek, A. Fix, and S. Houweling, "Space-borne remote sensing of $CO_2$, $CH_4$, and $N_2O$ by integrated path differential absorption lidar: a sensitivity analysis," Appl. Phys. B **90**, 593-608 (2008).
28. M. J. Kavaya, J. Y. Beyon, G. J. Koch, M. Petros, P. J. Petzar, U. N. Singh, B. C. Trieu, and J. Yu, "The Doppler aerosol wind (DAWN) airborne, wind-profiling coherent-detection Lidar system: overview and preliminary flight results," J. Atmos. Oceanic Technol. **31**, 826-842 (2014).
29. P. J. Roberts, F. Couny, H. Sabert, B. J. Mangan, D. P. Williams, L. Farr, M. W. Mason, A. Tomlinson, T. A. Birks, J. C. Knight, and P. St. J. Russell, "Loss in solid-core photonic crystal fibers due tointerface roughness scattering," Opt. Express **13**, 7779-7793 (2005).
30. K. R. Parameswaran, R. K. Route, J. R. Kurz, R. V. Roussev, M. M. Fejer, and M. Fujimura, "Highly efficient second-harmonic generation in buried waveguides formed by annealed and reverse proton exchange in periodically poled lithium niobate," Opt. Lett. **27**, 179-181 (2002).
31. R. V. Roussev, C. Langrock, J. R. Kurz, and M. M. Fejer, "Periodically poled lithium niobate waveguide sum-frequency generator for efficient single-photon detection at communication wavelengths," Opt. Lett. **29**, 1518-1520 (2004).
32. R. V. Roussev, "Optical-frequency mixers in periodically poled lithium niobate: materials, modeling and characterization," Ph. D. thesis (Department of Applied Physics, Stanford University, 2006).
33. A. Gröne and S. Kapphan, "Direct OH and OD librational absorption bands in $LiNbO_3$," J. of Phys. and Chem. of Solids **57**, 325-331 (1996).
34. D. H. Yoon, W. S. Yang, J. M. Kim, H. D. Yoon, Y. S. Kwak, J. S. Park, and H. Y. Lee, "Fabrication and properties of a 4×4 $LiNbO_3$ optical matrix switch," Mater. Trans. **43**, 1061-1064 (2002).
35. A. Kumar and S. Aditya, "Performance of S-bends for integrated-optic waveguides," Microwave Opt. Technol. Lett. **19**, 289-292 (1998).
36. T. C. Sum, A. A. Bettiol, S. Venugopal Rao, J. A. van Kan, A. Ramam, and F. Watt, "Proton beam writing of passive polymer optical waveguides," Proc. SPIE **5347**, 160-169 (2007).
37. H. A. Haus, W. P. Huang, S. Kawakami, and N. A. Whitaker, "Coupled-mode theory of optical waveguides," J. Lightwave Technol. **LT-5**, 16-23 (1987).
38. C. Langrock, "Classical and low-light-level detection and pulse characterization using optical-frequency mixers," Ph. D. thesis (Department of Electrical Engineering, Stanford University, 2007).
39. M. Zheng, G. Shentu, F. Ma, F. Zhou, H. Zhang, Y. Dai, X. Xie, Q. Zhang, and J. Pan, "Integrated four-channel all-fiber up-conversion single-photon-detector with adjustable efficiency and dark count," Rev. Sci. Instrum. **87**, 093115 (2016).

Table 1. Losses of the waveguide components for 1550 nm and 1950 nm

| Component | 1550 nm loss (dB) | 1950 nm loss (dB) |
| --- | --- | --- |
| Mode filter | 0.95 | 1.3 |
| Propagation loss | 0.6 | 0.6 |
| Directional coupler | 0.5 | 0.22 |
| Total | 2.05 | 2.12 |

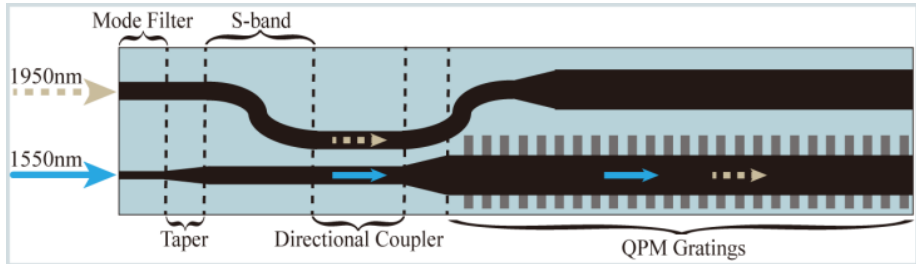

Fig. 1. Schematic of the integrated PPLN waveguides. The gray (dashed line) and blue (solid line) arrows represent 1950 nm and 1550 nm waves respectively.

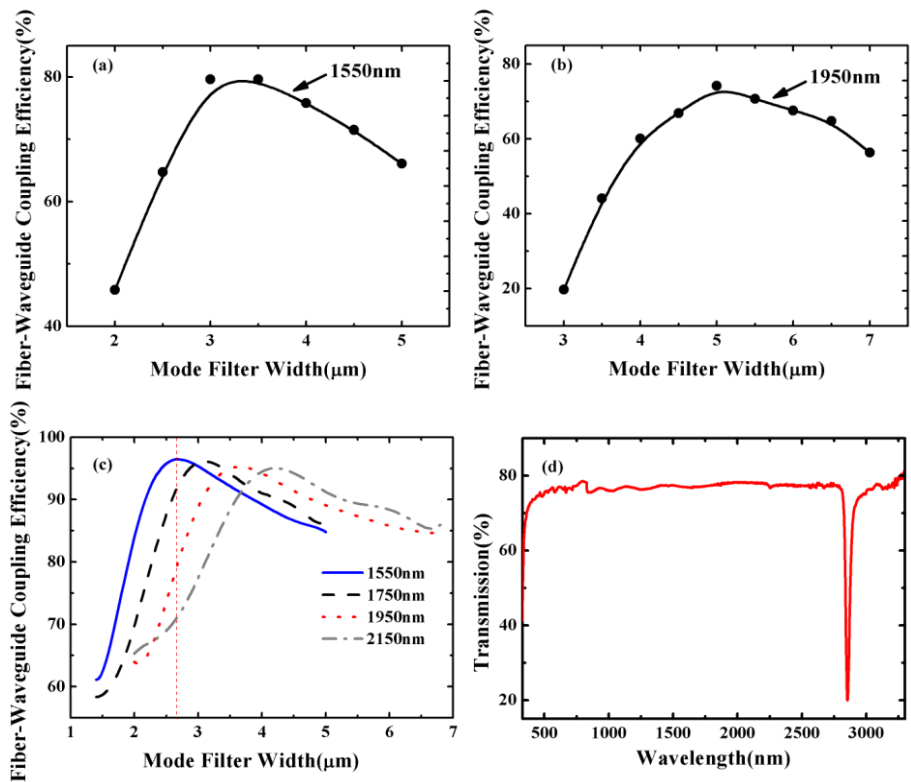

Fig. 2. The fiber-waveguide coupling efficiencies of (a) 1550 nm and (b) 1950 nm versus different mode filter widths (the solid lines are guidelines); (c) Simulation results of fiber-waveguide coupling efficiencies versus different mode filter widths at different wavelengths; (d) Transmission spectrum of a proton exchanged lithium niobate chip tested with a UV3600 spectrometer. The remarkable transmission loss in the flat region of the curve is caused by Fresnel reflections.

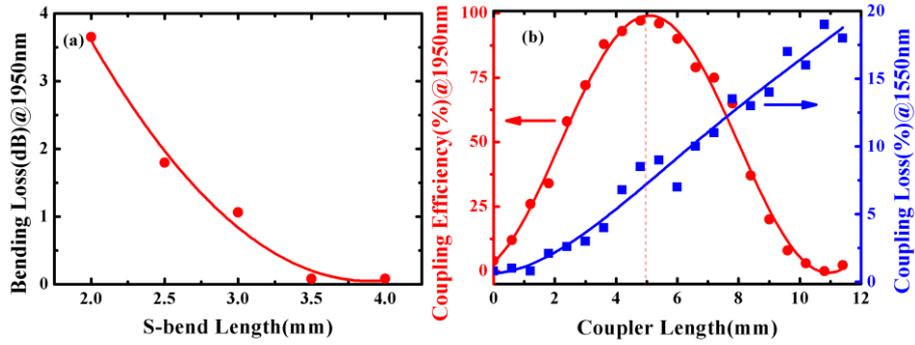

Fig. 3. (a) Bending loss of 1950 nm versus S-bend length in a cosine structure (the solid line is guideline); (b) 1950 nm coupling efficiency (solid-round dots) and 1550 nm coupling loss (solid-square dots) of directional coupler versus coupler length (the solid lines represent the fitting results [38]).

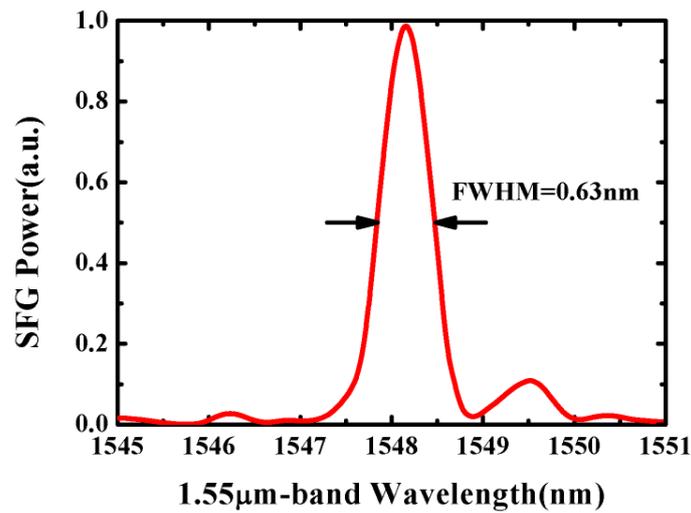

Fig. 4. Tuning curve for 1.55 μm band up-conversion process.

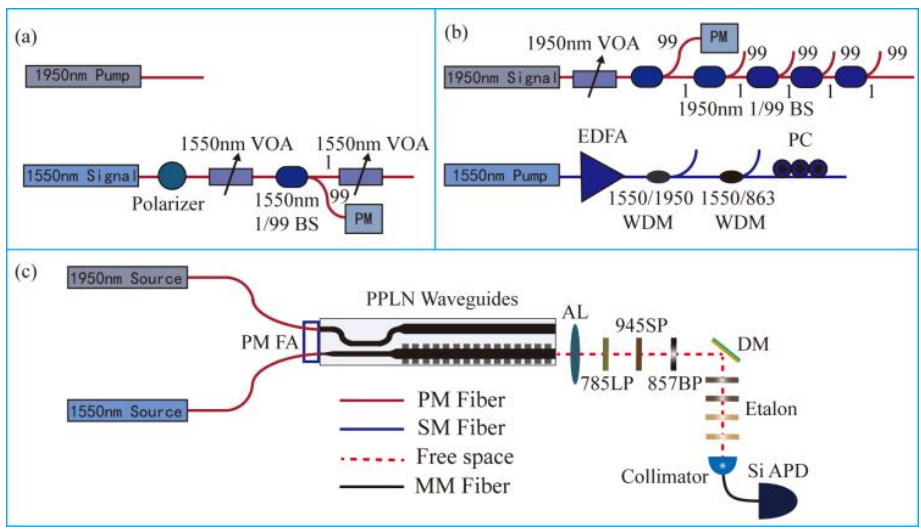

Fig. 5. (a) 1950 nm pump and single-photon-level 1550 nm signal; (b) Single-photon-level 1950 nm signal and 1550 nm pump; (c) Schematics of up-conversion single-photon detectors based on the integrated PPLN waveguides. VOA, variable optical attenuator; BS, beam splitter; PM, power meter; EDFA, erbium-doped fiber

amplifier; WDM, wavelength division multiplexer; PC, polarization controllers; PM FA, polarization maintaining fiber array; PPLN, periodically poled lithium niobate; AL, aspheric lens; LP, long-pass; SP, short-pass; BP, band-pass; DM, dichroic mirror; Si APD, silicon avalanche photodiode; PM fiber, polarization maintaining fiber; SM fiber, single-mode fiber; MM fiber, multimode fiber.

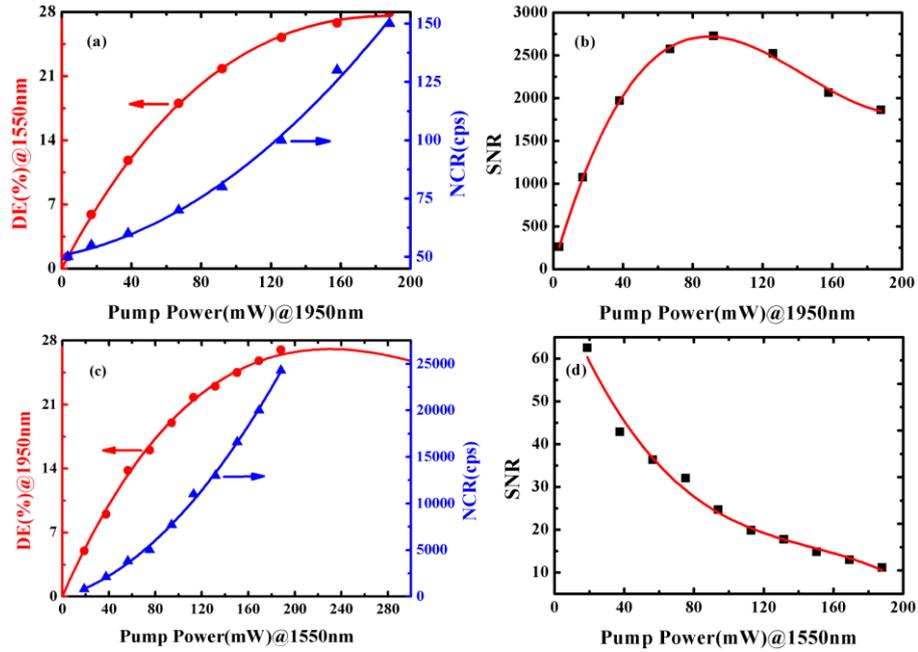

Fig. 6. DE (solid-round dots) and NCR (solid-triangle dots) versus pump power for the photon detection of (a) 1550 nm and (c) 1950 nm; SNR (solid-square dots) versus pump power for the photon detection of (b) 1550 nm and (d) 1950 nm (the solid lines are guidelines).